\documentclass{article}%
\usepackage{amsfonts}
\usepackage{amsmath}
\usepackage{amssymb}
\usepackage{graphicx}%
\providecommand{\U}[1]{\protect\rule{.1in}{.1in}}
\begin{document}

\title{Measuring the speed of quantum particles without a round-trip under
non-synchronized quantum clocks}
\author{Tomer Shushi\\Center for Quantum Science and Technology\\\& Department of Business Administration,\\Guilford Glazer Faculty of Business and Management,\\Ben-Gurion University of the Negev, Beer-Sheva, Israel\\ Email Address: tomershu@bgu.ac.il}
\maketitle

\begin{abstract}
One of the main issues in measuring the speed of light when it only travels
from one spatial position into another position, known as the one-way speed of
light, is that the clocks belonging to each separated spatial position are not
and, in principle, cannot be synchronized with sufficient precision. This
issue is the main reason why all of the measurements of the speed of light
until now have measured the two-way speed of light, i.e., measuring the speed
of light that travels from a source to another location and back to the
source, and so there is a need for only one clock to measure the speed. Here,
we show that it is possible, in principle, to measure the velocity of
particles that travel at the speed of light without assuming a round-trip once
we adopt a quantum mechanical description under two boundary conditions to the
state of the quantum system followed by the two-state-vector formalism while
assuming non-synchronized quantum clocks with unknown time dilation. We show
that the weak value of velocity can be measured for a test particle that has a
clock that is not synchronized with the clock of the quantum particle.
Following the proposed setup, when the weak value of the velocity is known
even without knowing the time states of the system, such a weak velocity is
the two-way speed of light. Otherwise, one has to impose assumptions regarding
the time states of the quantum clocks, which then give weak velocities that
can be slower or even faster than the two-way speed of light. We further
explore some fundamental implications of the setup. The proposed approach
opens a new avenue toward measuring the velocities of quantum particles while
overcoming relativistic issues regarding the synchronization of clocks.

\textit{Keywords:} quantum clocks, quantum foundations, relativistic
causality, time dilation, two-state-vector-formalism

\end{abstract}

\section{Introduction}

The speed of light is conventionally measured as a two-way speed, meaning that
light is sent from a source to another location and back to the source,
allowing for the round-trip time to be divided by the total distance traveled
[1-5]. The one-way speed of light means that we measure the speed of light
that only travels from a source to another location without a round-trip. This
would require perfectly synchronized clocks at both the departure and arrival
points. While there are no theoretical restrictions for measuring the one-way
speed of light, no experiment currently succeeds in measuring the one-way
speed of light. Various experiments have been claimed to measure the one-way
speed of light [6-8]. However, all of these experiments have detected the
two-way speed of light using different setups. For instance, in a 2009 paper
[7], the authors proposed a new way for such a measurement. However, only a
year after publishing their paper in [8], it has been shown that such
experiments do not measure the one-way speed of light but, in fact, the
two-way speed of light, and the problem remained open.

According to Einstein's theory of relativity, the synchronization of clocks
depends on the relative motion of the observers and the frames of reference in
which the measurements are taken. Any attempt to synchronize clocks across
different locations without taking into account these relativistic effects
would introduce errors, making the measurement unreliable. In the original
paper of Einstein proposing special relativity in 1905 [9], Einstein
postulated that the speed of light from spatial point $A$ to point $B$ is the
same from point $B$ to point $A,$ following the quote from his paper: "We have
not defined a common \textquotedblleft time\textquotedblright\ for $A$ and
$B$, for the latter cannot be defined at all unless we establish by
\textit{definition }that the \textquotedblleft time\textquotedblright%
\ required by light to travel from $A$ to $B$ equals the \textquotedblleft
time\textquotedblright\ it requires to travel from $B$ to $A$". While this
postulate seems pleasurable, there is currently no evidence that this is
indeed the case, i.e., that the speed of light is the same for every direction
in space. Every experiment that aims to measure the speed of light takes the
following general scheme:%

\includegraphics[width = \textwidth]{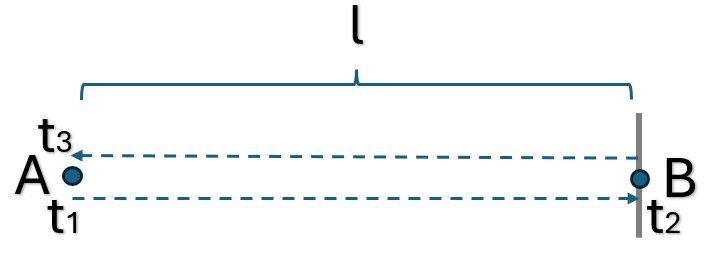}

Figure 1. The process for measuring the speed of light, starting from the
spatial position $A$ at time $t_{1}$ to separated spatial position $B$ (a
mirror) at time $t_{2}$ and back to the source at time $t_{3}$.

Then, following Einstein's original idea, the speed of light is calculated by%
\begin{equation}
c=\frac{2AB}{t_{3}-t_{1}}, \label{01}%
\end{equation}
i.e., it is the time that took the light to travel from point $A$ to $B$ and
back to $A.$ This is known as the \textit{Einstein synchronization}. Following
Einstein's original paper, the time of arrival of the light follows the
equation $t_{2}=t_{1}+\left(  t_{3}-t_{1}\right)  /2.$ As shown
by\ Reichenbach [1], we can consider an arbitrary rule for the time of arrival
followed by $t_{2}=t_{1}+\epsilon\left(  t_{3}-t_{1}\right)  $ for some
$0<\epsilon<1.$ This essentially corresponds to the speed of light in each
direction%
\begin{equation}
c_{\rightarrow}=\frac{1}{2}\frac{c}{\epsilon}, \label{c1}%
\end{equation}
and%
\begin{equation}
c_{\leftarrow}=\frac{1}{2}\frac{c}{1-\epsilon}, \label{c2}%
\end{equation}
so, in principle, it is possible that the speed of light in one direction does
not equal the speed of light in the opposite direction, i.e., $c_{\rightarrow
}\neq c_{\leftarrow}.$ This may be interpreted such that\ the one-way speed of
light is not isotropic in space, in the sense that when light travels in one
direction, it possesses a different speed than when it travels in the opposite
direction. While $c_{\rightarrow},c_{\leftarrow}$ can have different values,
they are mutually dependent. In particular, they obey the equation%
\begin{equation}
\frac{c_{\rightarrow}\cdot\left(  t_{2}-t_{1}\right)  +c_{\leftarrow}%
\cdot\left(  t_{3}-t_{2}\right)  }{t_{3}-t_{1}}=\frac{2l}{t_{3}-t_{1}}=c.
\label{02}%
\end{equation}
If $c_{\rightarrow}$ is faster/slower than the two-way speed of light $c,$
$c_{\leftarrow}$ compensates it with slower/faster speed, such that the
average of $c_{\rightarrow}$ and $c_{\leftarrow}$ describes the usual
(two-way) speed of light.

In the following, we demonstrate how the problem of measuring particle
velocity without assuming a round-trip is resolved by adopting a quantum
mechanical description based on the two-state-vector formalism (TSVF). In the
proposed setup, we show that the weak value of velocity can be measured for a
test particle with a clock that is not synchronized with the clock of the
quantum particle, even when the time dilation between the clocks is entirely
unknown. We examine some basic implications and\ show that the proposed
setup\ does not break relativistic causality, and also examine the case of
position-dependent speed of light.

\section{Results}

Before we introduce the proposed setup, we review the basic elements in
consideration of time as a dynamical quantum variable and not a mere
parameter. Let us consider a\ quantum system that contains\ a clock $B$ and
the rest of the system denoted by $R,$ where each of the sub-systems is
described by\ vectors in the Hilbert spaces $\mathcal{H}_{A}$ and
$\mathcal{H}_{R},$ respectively. The entire system is then described by the
quantum state\ $\left\vert \left.  \Psi\right\rangle \right\rangle
\in\mathcal{H}_{B}\otimes\mathcal{H}_{R}.$ Assuming that we have a closed
quantum system, it follows the\ Wheeler-DeWitt equation%
\begin{equation}
\left(  H_{B}+H_{R}\right)  \left\vert \left.  \Psi\right\rangle \right\rangle
=0, \label{03}%
\end{equation}
which suggests we have a\ (stationary) state of the system that does not
change with respect to an external time, with a zero eigenvalue of energy (see
[10-18]). We can describe the time observable that is associated with clock
$B$ by proposing a time operator\ $T_{B}$ in the Hilbert space $\mathcal{H}%
_{B},$ which is observable in units of time. We then define the time
eigenvectors, $\left\vert t_{B}\right\rangle $, and time eigenvalues, $t_{B}$,
of $T_{B}$ as the solutions of the equation $T_{B}\left\vert t_{B}%
\right\rangle =t_{B}\left\vert t_{B}\right\rangle $. Similar to a standard
quantum system, we follow the usual commutation relation $\left[  T_{B}%
,H_{B}\right]  =i\hbar I,$ where the Hamiltonian of the clock is
$H_{B}=-i\hbar\partial/\partial t_{B}$ (see, again, [12-13]).$\ $The unitary
evolution of our time states is then given by
\begin{equation}
\left\vert t_{0}+t_{B}\right\rangle =e^{-iH_{B}t_{B}/\hbar}\left\vert
t_{0}\right\rangle , \label{04}%
\end{equation}
for an initial time state $\left\vert t_{0}\right\rangle .$ Similar to [13],
we define an interaction\ $H_{int}\left(  T_{B}\right)  ,$ which is\ a
time-dependent term of the evolution of system $R$ set by clock $B,$ and it
quantifies the interaction between the clock $B$ and the rest of the system
$R.$ Then, the total Hamiltonian of our quantum system is given by
$H_{total}=H_{B}+H_{R}+H_{int}\left(  T_{B}\right)  .$ Following the
Wheeler-DeWitt equation, we have $H_{total}\left\vert \left.  \Psi
\right\rangle \right\rangle =0,$ and by applying the time eigenstate
$\left\vert t_{B}\right\rangle $ of $T_{B}$ on the left, we can obtain the
time-dependent Schr\"{o}dinger equation for the quantum state $\left\vert
\psi\left(  t_{B}\right)  \right\rangle :=\left.  \left\langle t_{B}%
|\Psi\right\rangle \right\rangle ,$
\begin{equation}
i\hbar\frac{\partial}{\partial t_{B}}\left\vert \psi\left(  t_{B}\right)
\right\rangle =\left[  H_{R}+H_{int}\left(  t_{B}\right)  \right]  \left\vert
\psi\left(  t_{B}\right)  \right\rangle , \label{05}%
\end{equation}
where the equation provides us the\ evolution of the wavefunction of system
$R$ with respect to clock $B$ (see, [14-15,18]).

Let us now consider a system containing two synchronized clocks, $A$ and $B$,
where $A$ is the internal clock of the rest of the system, i.e.,\ $R=A+S.$ The
Hamiltonian of the system is then $H_{total}=H_{A}+H_{S}+H_{B}+H_{int}\left(
T_{B}\right)  $, and the Schr\"{o}dinger equation is $i\hbar\frac{\partial
}{\partial t_{B}}\left\vert \psi\left(  t_{B}\right)  \right\rangle =\left[
H_{A}+H_{S}+H_{int}\left(  T_{B}\right)  \right]  \left\vert \psi\left(
t_{B}\right)  \right\rangle .$

A key element in the one-way speed of light problem is the\ break of
synchronicity between the clocks. We can model such a break by adding to the
Hamiltonian the term
\begin{equation}
g\left(  T_{B}\right)  H_{A}, \label{06}%
\end{equation}
where $g\left(  u\right)  ,$ $u\geq0,$ is some positive\ function, where
$\int_{0}^{\infty}g\left(  t\right)  dt>0$. Then, following [12-13],
our\ Schr\"{o}dinger equation is given by%
\begin{equation}
i\hbar\frac{\partial}{\partial t_{B}}\left\vert \psi\left(  t_{B}\right)
\right\rangle =\left(  I+g\left(  t_{B}\right)  \right)  H_{A}\left\vert
\psi\left(  t_{B}\right)  \right\rangle . \label{07}%
\end{equation}
The\ Heisenberg picture allows us to get the exact break of synchronicity
between clocks $A$ and $B,$ followed by the time observable of clock $A,$
$T_{A},$ with respect to the ticks of clock $B,$%
\begin{equation}
\frac{d}{dt_{B}}T_{A}=-\frac{i}{\hbar}\left[  T_{A},H_{R}+g\left(
t_{B}\right)  H_{A}\right]  =I+g\left(  t_{B}\right)  . \label{breaksyn1}%
\end{equation}
We note that when $g$ vanishes, the flow of time in both clocks is the same,
and in case $g\neq0,$ we see that clock $A$ ticks faster than clock $B$.

We can now propose the setup demonstrating how to measure the velocity of
quantum particles without a round-trip using the TSVF. For the setup of the
weak measurement of velocity, we adopt the framework proposed in [19], which
originally provided a way to obtain Chernekov radiation in a vacuum based on
the TSVF and the weak velocity of the quantum particles.

TSVF proposes a time-symmetric picture for the evolution of quantum systems
based on two states: a state that is prepared and evolves unitarily forward
from the past and a post-selected state. It has been shown that the TSVF
allows for the measurement between the pre- and post-selected states, such
that the quantum state of the particle is not disturbed, and so hidden
information and odd quantum effects of the quantum particle are detected [20-26].

We consider a system that contains a quantum particle $S$ with an internal
clock $A$ and an external clock $B,$ followed by the total Hamiltonian%
\begin{equation}
H_{Tot}=H_{A}+H_{S}+H_{B}+H_{int}\left(  T_{B}\right)  +g\left(  T_{B}\right)
H_{A}, \label{08}%
\end{equation}
where $H_{A}/H_{B}$ is the (time) Hamiltonian of the quantum clock that
generates the translations in time $T_{A}/T_{B}$. $H_{S}$ is the Hamiltonian
of the quantum particle and is given by%
\begin{equation}
H_{S}=p_{z}v_{z}, \label{H_vz1}%
\end{equation}
where $p_{z}=-i\hbar\partial/\partial z,$ and%
\begin{equation}
v_{z}=c\cdot\sigma_{z} \label{09}%
\end{equation}
acts on the internal Hilbert space of the particle, where $c$ is the speed of
light, and $N>0$ is some integer. The Pauli matrix $\sigma_{z}$ operates on
the internal Hilbert space. We note that it does not represent spin, and so
the particle has no electric or magnetic dipole moment. The eigenvalues of the
velocity along the $z$ direction $v_{z}$ are $-c,+c,$ and so if the only
allowed values of $v_{z}$ are its eigenvalues, the quantum particle can only
travel at the speed of light $c$.

The particle moves with velocity $v_{z}$ in the $z$ direction, i.e., which can
be described in the Heisenberg picture, $\frac{d}{dt_{A}}x=-\frac{i}{\hbar
}\left[  x,H_{S}\right]  =0,~\frac{d}{dt_{A}}y=-\frac{i}{\hbar}\left[
y,H_{S}\right]  =0,$\ and%
\begin{equation}
\frac{d}{dt_{A}}z=-\frac{i}{\hbar}\left[  z,H_{S}\right]  =v_{z}. \label{011}%
\end{equation}
We assume that the system is initially prepared in the form%
\begin{equation}
\left\vert \left.  \Psi_{in}\right\rangle \right\rangle =\left\vert \Psi
_{in}^{S},\mathcal{T}_{in}\right\rangle \Phi\left(  \boldsymbol{x},0\right)  ,
\label{012}%
\end{equation}
where the quantum particle is prepared in the pre-selected state%
\begin{equation}
\left\vert \Psi_{in}^{S},\mathcal{T}_{in}\right\rangle =\frac{1}{\sqrt{2}%
}\left(  \left\vert \uparrow\right\rangle \left\vert \mathcal{T}_{in}%
^{+}\right\rangle +\left\vert \downarrow\right\rangle \left\vert
\mathcal{T}_{in}^{-}\right\rangle \right)  , \label{013}%
\end{equation}
for initial time states $\left\vert \mathcal{T}_{in}^{-/+}\right\rangle ,$ and
$\Phi$ describes an additional (test)\ particle that is approximately
localized in the origin, $\boldsymbol{x}=\left(  x,y,z\right)  =0,$ and it
takes the Gaussian form $\Phi\left(  \boldsymbol{x},0\right)  =\left(
\varepsilon^{2}\pi\right)  ^{-3/4}e^{-\boldsymbol{x}^{T}\boldsymbol{x}%
/2\varepsilon^{2}}$ for $\varepsilon>0$ which gives the dispersion of the
Gaussian, and it evolves according to clock $B$.

We post- select the quantum particle and the clocks' states and find it in the
state%
\begin{equation}
\left\vert \Psi_{fin}^{S},\mathcal{T}_{fin}\right\rangle =\alpha\left\vert
\uparrow\right\rangle \left\vert \mathcal{T}_{fin}^{+}\right\rangle
+\beta\left\vert \downarrow\right\rangle \left\vert \mathcal{T}_{fin}%
^{-}\right\rangle , \label{014}%
\end{equation}
for (known) real-valued coefficients $\alpha,\beta,$ such that $\alpha
^{2}+\beta^{2}=1,$ and final time states $\left\vert \mathcal{T}_{fin}%
^{-/+}\right\rangle .$ Since we do not have any knowledge about the time
dilation between the clocks, i.e., about $g,$ we assume that we also do not
have knowledge about the time states $\left\vert \mathcal{T}_{in/fin}%
^{-/+}\right\rangle ,$ we do, however, assume that they are normalized similar
to any other standard quantum state and that $\left\langle \mathcal{T}%
_{fin}^{-/+}|\mathcal{T}_{in}^{-/+}\right\rangle \neq0$.

The function $g$ generates the break in synchronicity between the clocks, as
shown in (\ref{breaksyn1}), implying a break in their synchronicity, as
illustrated by (\ref{breaksyn1})%
\begin{equation}
\frac{d}{dt_{B}}T_{A}=-\frac{i}{\hbar}\left[  T_{A},H_{A}+H_{S}+H_{B}%
+H_{int}\left(  T_{B}\right)  +g\left(  T_{B}\right)  H_{A}\right]
=I+g\left(  t_{B}\right)  , \label{016}%
\end{equation}
where the internal clock of the particle, clock $A\,,$ ticks faster than clock
$B,$ the clock of the test particle.

Similar to the setup proposed in [19], $\Phi\left(  \boldsymbol{x}%
,t_{B}\right)  $ provides the quantum state of the test particle with respect
to clock $B,$ and is given by%
\begin{align}
\Phi\left(  \boldsymbol{x},t_{B}\right)   &  =\left\langle \Psi_{fin}%
^{S},\mathcal{T}_{fin}\right\vert e^{-iH_{Tot}t_{B}/\hbar}\left\vert \Psi
_{in}^{S},\mathcal{T}_{in}\right\rangle \Phi\left(  \boldsymbol{x},0\right)
\label{017}\\
&  =\left\langle \Psi_{fin}^{S},\mathcal{T}_{fin}\right\vert e^{-i\left(
H_{\text{clocks}}+g\left(  t_{B}\right)  H_{A}+p_{z}v_{z}\right)  t_{B}/\hbar
}\left\vert \Psi_{in}^{S},\mathcal{T}_{in}\right\rangle \Phi\left(
\boldsymbol{x},0\right) \nonumber
\end{align}
where $H_{\text{clocks}}:=H_{A}+H_{B}+H_{int}\left(  T_{B}\right)  .$ We
recall that $g$ is unknown since we do not know the time dilation between
clocks $A$ and $B.$

Then, for a short enough time $t_{B}$ and under the condition%
\[
g\left(  t_{B}\right)  <<\frac{1}{t_{B}},
\]
we have%
\begin{align}
\Phi\left(  \boldsymbol{x},t_{B}\right)   &  \approx\left\langle \Psi
_{fin}^{S},\mathcal{T}_{fin}\right\vert 1-\frac{i}{\hbar}\left(
H_{\text{clocks}}+g\left(  t_{B}\right)  H_{A}+p_{z}v_{z}\right)
t_{B}\left\vert \Psi_{in}^{S},\mathcal{T}_{in}\right\rangle \Phi\left(
\boldsymbol{x},0\right) \nonumber\\
&  =\left\langle \Psi_{fin}^{S},\mathcal{T}_{fin}\right\vert 1-\frac{i}{\hbar
}\left(  \left\langle H_{\text{clocks}}\right\rangle _{w}+g\left(
t_{B}\right)  \left\langle H_{A}\right\rangle _{w}+p_{z}\left\langle
v_{z}\right\rangle _{w}\right)  t_{B}\left\vert \Psi_{in}^{S},\mathcal{T}%
_{in}\right\rangle \Phi\left(  \boldsymbol{x},0\right) \nonumber\\
&  \approx\kappa\left\langle \Psi_{fin}^{S},\mathcal{T}_{fin}\right\vert
e^{-ip_{z}\left\langle v_{z}\right\rangle _{w}t_{B}/\hbar}\left\vert \Psi
_{in}^{S},\mathcal{T}_{in}\right\rangle \Phi\left(  \boldsymbol{x},0\right)
\nonumber\\
&  =\kappa\left\langle \Psi_{fin}^{S},\mathcal{T}_{fin}|\Psi_{in}%
^{S},\mathcal{T}_{in}\right\rangle \cdot\Phi\left(  x,y,z-\left\langle
v_{z}\right\rangle _{w}t_{B},0\right)  , \label{phi_dis1}%
\end{align}
where $\kappa=e^{-i\left(  \left\langle H_{\text{clocks}}\right\rangle
_{w}+g\left(  t_{B}\right)  \left\langle H_{A}\right\rangle _{w}\right)
t_{B}/\hbar},$ and $\left\langle A\right\rangle _{w}$ is the weak-value of an
observable $A,$
\begin{equation}
\left\langle A\right\rangle _{w}:=\frac{\left\langle \Psi_{fin}^{S}%
,\mathcal{T}_{fin}\right\vert A\left\vert \Psi_{in}^{S},\mathcal{T}%
_{in}\right\rangle }{\left\langle \Psi_{fin}^{S},\mathcal{T}_{fin}|\Psi
_{in}^{S},\mathcal{T}_{in}\right\rangle }. \label{020}%
\end{equation}
Eq. (\ref{phi_dis1}) implies that the wavefunction of the test particle is
displaced by $\left\langle v_{z}\right\rangle _{w}t_{B}.$

The velocity of the quantum particle is directly related to the pre- and post-
selected states, followed by the weak velocity $\left\langle v_{z}%
\right\rangle _{w},$%
\begin{equation}
\left\langle v_{z}\right\rangle _{w}=\frac{\left(  \left\langle \mathcal{T}%
_{fin}^{+}\right\vert \left\langle \uparrow\right\vert \alpha+\left\langle
\mathcal{T}_{fin}^{-}\right\vert \left\langle \downarrow\right\vert
\beta\right)  v_{z}\left(  \left\vert \uparrow\right\rangle \left\vert
\mathcal{T}_{in}^{+}\right\rangle +\left\vert \downarrow\right\rangle
\left\vert \mathcal{T}_{in}^{-}\right\rangle \right)  }{\left(  \left\langle
\mathcal{T}_{fin}^{+}\right\vert \left\langle \uparrow\right\vert
\alpha+\left\langle \mathcal{T}_{fin}^{-}\right\vert \left\langle
\downarrow\right\vert \beta\right)  \left(  \left\vert \uparrow\right\rangle
\left\vert \mathcal{T}_{in}^{+}\right\rangle +\left\vert \downarrow
\right\rangle \left\vert \mathcal{T}_{in}^{-}\right\rangle \right)  }%
=\frac{\left\langle \mathcal{T}_{fin}^{+}|\mathcal{T}_{in}^{+}\right\rangle
\alpha-\left\langle \mathcal{T}_{fin}^{-}|\mathcal{T}_{in}^{-}\right\rangle
\beta}{\left\langle \mathcal{T}_{fin}^{+}|\mathcal{T}_{in}^{+}\right\rangle
\alpha+\left\langle \mathcal{T}_{fin}^{-}|\mathcal{T}_{in}^{-}\right\rangle
\beta}\cdot c, \label{021}%
\end{equation}
which can, in principle, take different values than the eigenvalues of
$v_{z},$ i.e., the two-way speed of light.

\bigskip

The following Figure illustrates the proposed setup of the gedanken experiment.%

\includegraphics[width = \textwidth]{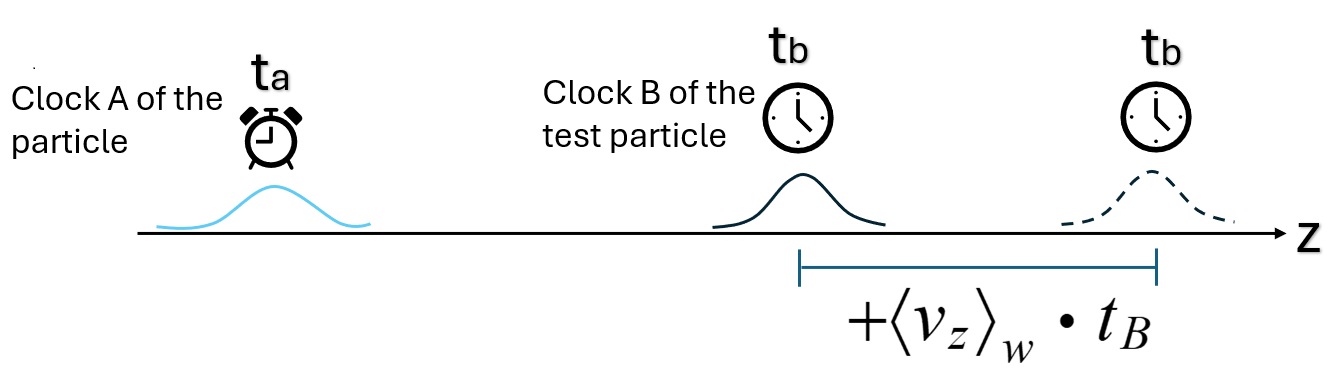}

Figure 2. A scheme of the proposed setup. The experiment is conducted along
the $z$-axis, where the quantum particle with internal clock $A$ is pre- and
post- selected followed by a test particle with internal clock $B.$ The test
particle is then shifted by $\left\langle v_{z}\right\rangle _{w}\cdot t_{B}$
in the $z$-axis.

It is important to emphasize that, unlike the expected value of velocity,
$\left\langle v_{z}\right\rangle $, which is a statistical quantity of the
system, the weak values are actual physical quantities, and so the measurement
that gives $\left\langle v_{z}\right\rangle _{w}$ essentially gives the
velocity of the quantum particle under the specified pre- and post- selected
states of the quantum system. We note that we have an isotropy of the speed of
light (of the particle) only statistically, following the expectation value of
$v_{z},$ $\left\langle v_{z}\right\rangle =0,$ and the eigenvalues $-c,c.$

\bigskip We can define a quantity, $\tau$, for the ratio between the time
amplitudes%
\begin{equation}
\tau=\frac{\left\langle \mathcal{T}_{fin}^{-}|\mathcal{T}_{in}^{-}%
\right\rangle }{\left\langle \mathcal{T}_{fin}^{+}|\mathcal{T}_{in}%
^{+}\right\rangle }, \label{tau01}%
\end{equation}
leading to the weak velocity of the form%
\begin{equation}
\left\langle v_{z}\right\rangle _{w}=\frac{\alpha-\tau\beta}{\alpha+\tau\beta
}\cdot c. \label{v_tau1}%
\end{equation}
Here, $\tau$ is the unknown parameter, since we do not have the knowledge
about the time states, and so we also do not have the knowledge about the
transition amplitudes $\left\langle \mathcal{T}_{fin}^{-/+}|\mathcal{T}%
_{in}^{-/+}\right\rangle .$ We can, however, find the value of $\left\langle
v_{z}\right\rangle _{w}$ in specific cases. In the case where $\beta=0$, the
weak velocity becomes the speed of light,
\begin{equation}
\left\langle v_{z}\right\rangle _{w}=c. \label{v_equal_c1}%
\end{equation}
The proposed setup reveals that when $\left\langle v_{z}\right\rangle _{w}$ is
known and is independent on the time states, the speed is the two-way speed of
light. When the initial and final times states are the same, i.e., $\left\vert
\mathcal{T}_{in}^{-/+}\right\rangle =\left\vert \mathcal{T}_{fin}%
^{-/+}\right\rangle ,$ from the orthonormality property of quantum states we
have $\left\langle \mathcal{T}_{fin}^{-/+}|\mathcal{T}_{in}^{-/+}\right\rangle
=\left\langle \mathcal{T}_{in}^{-/+}|\mathcal{T}_{in}^{-/+}\right\rangle =1,$
and the weak velocity also becomes a known quantity with by having $\tau
\equiv1,$ and a potential violation of the two-way speed of light, with
$\left\langle v_{z}\right\rangle _{w}=\frac{\alpha-\beta}{\alpha+\beta}\cdot
c.$

Unlike standard (strong) measurements of the velocity, where the measurements
only give the eigenvalues of $v_{z},$ weak values can give values out of the
spectrum of the eigenvalues. We considered a particle that can only travel at
the speed of light, with eigenvalues $-c,c$ along the $z$-axis; however, the
weak velocity $\left\langle v_{z}\right\rangle _{w}$ can be slower and even
faster than $c.$ We can now relate the $\epsilon$ in the model of the one-way
speed of light with the weak velocity, following formula (\ref{c1}), with
having%
\begin{equation}
\epsilon=\frac{1}{2}\frac{\alpha+\tau\beta}{\alpha-\tau\beta} \label{022}%
\end{equation}
and since $0<\epsilon<1,$ we assume that $\tau$ is real-valued, and we are
restricted to the coefficients of the post- selected state that satisfies%
\begin{equation}
\tau<\frac{\alpha}{3\beta}. \label{023}%
\end{equation}

The proposed setup can also be used for particles that have a range of
possible speeds bounded by the speed of light. In particular, instead of only
eigenvalues $-c,c$ of $v_{z}$ we can consider the case where the eigenvalues
of the velocity along the $z$ direction $v_{z}$ are $-c,-c+2c/N,...,c-2c/N,c.$
This corresponds with the velocity observable
\begin{equation}
v_{z}=\frac{c}{N}\sum_{i=1}^{N}\sigma_{z}^{\left(  i\right)  }, \label{024}%
\end{equation}
where, again,\ the Pauli matrices $\sigma_{z}^{\left(  i\right)  }$ operate on
the internal Hilbert space. Then, we can obtain the same experiment as the
previous one while considering the pre- and post- selected states of the
quantum system%
\begin{equation}
\left\vert \Psi_{in}^{S},\mathcal{T}_{in}\right\rangle =2^{-N/2}\otimes
_{i=1}^{N}\left(  \left\vert \uparrow_{i}\right\rangle \left\vert
\mathcal{T}_{in}^{+}\right\rangle +\left\vert \downarrow_{i}\right\rangle
\left\vert \mathcal{T}_{in}^{-}\right\rangle \right)  , \label{025}%
\end{equation}
and%
\begin{equation}
\left\vert \Psi_{fin}^{S},\mathcal{T}_{fin}\right\rangle =\otimes_{i=1}%
^{N}\left(  \alpha\left\vert \uparrow_{i}\right\rangle \left\vert
\mathcal{T}_{fin}^{+}\right\rangle +\beta\left\vert \downarrow_{i}%
\right\rangle \left\vert \mathcal{T}_{fin}^{-}\right\rangle \right)  ,
\label{026}%
\end{equation}
respectively, then similar to the original setup, the weak velocity is given
by $\left\langle v_{z}\right\rangle _{w}=\frac{\alpha-\tau\beta}{\alpha
+\tau\beta}\cdot c.$

\subsection{Relativistic causality}

The $\epsilon$ parameter in (\ref{c1}) and (\ref{c2}) implies a speed that is
higher than $c$ in at least one direction. However, it does not violate
relativistic causality, since we can only measure two-way speed of light in
the original classical setup. We provided a setup that can also exceed the
speed of light whenever%
\begin{equation}
\alpha\cdot\tau\beta<0, \label{027}%
\end{equation}
since then $\left\langle v_{z}\right\rangle _{w}>c.$ However, this also does
not violate relativistic causality, as mentioned in [19].

In the case
\begin{equation}
\left\vert \frac{\alpha-\tau\beta}{\alpha+\tau\beta}\right\vert <1,
\label{condi1}%
\end{equation}
i.e.
\begin{equation}
\left\vert \left\langle v_{z}\right\rangle _{w}\right\vert <c, \label{v_w_c1}%
\end{equation}
we do not have any problem with relativistic causality since the speed is less
than the speed of light. In the case where $\left\vert \left\langle
v_{z}\right\rangle _{w}\right\vert >c,$ we may have a problem with
relativistic causality. However, no violation of relativistic causality is
taking place. The reason is that since $\Phi$ is an analytic function with
respect to $z,$ we have a change in $\Phi$, $\Phi\left(  x,y,z-\left\langle
v_{z}\right\rangle _{w}t_{B},0\right)  ,$ which does not transmit any
information, because the same message is passed to all $\mathbf{x}$ and $t$
(see, [19]).$\ $Thus, there is no way in which relativistic causality can be violated.

\subsection{The case of position-dependent speed of light}

Let us now examine the case of position-dependent speed of light. Since the
beginning days of relativity theory, the position-dependent speed of light,
$c\left(  \boldsymbol{x}\right)  ,$ has been explored. Such consideration even
appeared in Einstein's paper in 1911 [27]. Such models of varying speed of
light (VSL), however, have not gained much attention in the literature. In
recent years, there has been a relative rise in exploring such hypothesized
possibilities under certain various conditions [28-31]. Such theories mainly
aim at exploring new cosmological models for exploring, for example, the
beginning of the universe. While the position-dependent speed of light is
still merely a theoretical, speculative theory, it may provide a new pathway
toward a possible theory that links concepts from quantum mechanics and
general relativity into a single coherent scheme. There are also various
attempts to establish experimental setups to test the possibility of
position-dependent speed of light (see, e.g., [28]). When considering the
position-dependent speed of light, one may also naturally consider other
extensions of physical constants that become variables for getting a
physically meaningful model of reality. Various models have been proposed, one
of which is the gravitational constant and the Planck constant, which depend
on the speed of light. Let us consider how $c\left(  \boldsymbol{x}\right)  $
and a reduced Planck constant that depends on $c\left(  \boldsymbol{x}\right)
,\hbar\left(  \boldsymbol{x}\right)  >0$ $\forall\boldsymbol{x},$ are
considered in the proposed model.

The standard momentum operator is given by $\widehat{p}_{j}=-i\hbar
_{0}\partial/\partial x_{j},j=1,2,3,$ where $\hbar_{0}$ is the (reduced)
Planck constant. When considering $\hbar$ to be variable, and in particular,
$\hbar\left(  \boldsymbol{x}\right)  ,$ there is no unique form for the
Hamiltonian of the system. While in the standard non-relativistic case, the
Hamiltonian is\ $H=\widehat{\boldsymbol{p}}^{2}/2m+V\left(  \boldsymbol{x}%
\right)  $ where $\widehat{\boldsymbol{p}}=\left(  \widehat{p}_{1}%
,\widehat{p}_{2},\widehat{p}_{3}\right)  .$ Here, since%
\begin{equation}
\left[  \hbar\left(  \boldsymbol{x}\right)  ,\widehat{p}_{j}\right]
\neq0,\text{\ \ }j=1,2,3, \label{h01}%
\end{equation}
there is no particular way of how to substitute $\hbar\left(  \boldsymbol{x}%
\right)  $ in the momentum operator.

Following [32,33], we define a general form for the Hamiltonian%
\begin{equation}
H_{S}=\Pi_{z,\hbar}v_{z}, \label{h02}%
\end{equation}
for the deformed momentum operator%
\begin{equation}
\Pi_{z,\hbar}=\sqrt{\hbar}\left(  -i\frac{\partial}{\partial_{z}}\right)
\sqrt{\hbar}. \label{h03}%
\end{equation}
We note that by adopting the deformed momentum operator, we also have a
deformed commutation relation,
\begin{equation}
\left[  z,\Pi_{z,\hbar}\right]  =i\hbar\left(  \boldsymbol{x}\right)  .
\label{h04}%
\end{equation}
Going back to our experiment, we have%

\begin{align}
\Phi\left(  \boldsymbol{x},t_{B}\right)   &  \approx\left\langle \Psi
_{fin}^{S},\mathcal{T}_{fin}\right\vert 1-\frac{i}{\hbar}\left(
H_{\text{clocks}}+g\left(  t_{B}\right)  H_{A}+\Pi_{z,\hbar}v_{z}\right)
t_{B}\left\vert \Psi_{in}^{S},\mathcal{T}_{in}\right\rangle \Phi\left(
\boldsymbol{x},0\right)  \nonumber\\
&  =\left\langle \Psi_{fin}^{S},\mathcal{T}_{fin}\right\vert 1-\frac{i}{\hbar
}\left(  \left\langle H_{\text{clocks}}\right\rangle _{w}+g\left(
t_{B}\right)  \left\langle H_{A}\right\rangle _{w}+p_{z}\left\langle
v_{z}\right\rangle _{w}\right)  t_{B}\left\vert \Psi_{in}^{S},\mathcal{T}%
_{in}\right\rangle \Phi\left(  \boldsymbol{x},0\right)  \nonumber\\
&  \approx\kappa\cdot\left\langle \Psi_{fin}^{S},\mathcal{T}_{fin}\right\vert
\exp\left\{  -i\frac{1}{\hbar\left(  \boldsymbol{x}\right)  }\Pi_{z,\hbar
}\left\langle v_{z}\right\rangle _{w}t_{B}\right\}  \left\vert \Psi_{in}%
^{S},\mathcal{T}_{in}\right\rangle \Phi\left(  \boldsymbol{x},0\right)
,\label{h05}%
\end{align}
where%
\begin{equation}
\left\langle v_{z}\right\rangle _{w}=\frac{\alpha-\tau\beta}{\alpha+\tau\beta
}\cdot c\left(  \boldsymbol{x}\right)  ,\label{h06}%
\end{equation}
and%
\begin{equation}
\frac{1}{\hbar\left(  c\left(  \boldsymbol{x}\right)  \right)  }\Pi_{z,\hbar
}\left\langle v_{z}\right\rangle _{w}=\frac{\alpha-\tau\beta}{\alpha+\tau
\beta}\frac{1}{\sqrt{\hbar\left(  c\left(  \boldsymbol{x}\right)  \right)  }%
}\left(  -i\frac{\partial}{\partial z}\right)  \sqrt{\hbar\left(
\boldsymbol{x}\right)  }\cdot c\left(  \boldsymbol{x}\right)  .\label{h07}%
\end{equation}
If we assume that the shift of $\Phi$ is similar to the case of constant speed
of light and Planck constant, then we have to assume that%
\begin{equation}
\sqrt{\hbar\left(  \boldsymbol{x}\right)  }\cdot c\left(  \boldsymbol{x}%
\right)  =\sqrt{\Lambda},\label{h08}%
\end{equation}
for some constant $\Lambda>0,$ and so we establish the relation%
\begin{equation}
\hbar\left(  \boldsymbol{x}\right)  =\frac{\Lambda}{c\left(  \boldsymbol{x}%
\right)  ^{2}},\label{h09}%
\end{equation}
implying that $\hbar$ is a function of $\boldsymbol{x}$ through $c\left(
\boldsymbol{x}\right)  .$ In this case, when we assume that $\Lambda
=\sqrt{\hbar_{0}}c_{0}$ where $\hbar_{0}$ is the original (reduced) Planck
constant and $c_{0}$ is the (two-way) speed of light in vacuum, we have%
\begin{equation}
\Phi\left(  x,y,z-\sqrt{\hbar_{0}/\hbar\left(  c\left(  \boldsymbol{x}\right)
\right)  }\left\langle v_{z}^{0}\right\rangle _{w}t_{B},0\right)  \label{h010}%
\end{equation}
where $\left\langle v_{z}^{0}\right\rangle _{w}=\frac{\alpha-\tau\beta}%
{\alpha+\tau\beta}c_{0}\ $is the weak-velocity in case the (two-way) speed of
light is a constant $c_{0}.$

\section{Discussion}

While all of the experiments for measuring the speed of light have measured
the two-way speed of light, there is still a quest for obtaining adequate
setups that allow the measurement without the consideration of a round-trip.
We have shown that it is possible, in principle, to measure the velocity of
particles that travel at the speed of light without assuming a round-trip once
we adopt a quantum mechanical description when considering two boundary
conditions to the quantum system followed by the two-state-vector formalism
while assuming non-synchronized quantum clocks with unknown time dilation.
Followed by the proposed setup, a test particle is shifted by the amount of
$\left\langle v_{z}\right\rangle _{w}t_{B}$, which allows finding the weak
velocity of the quantum particle, $\left\langle v_{z}\right\rangle _{w}.$ The
weak velocity takes, in general, different values than the usual eigenvalues
of the velocity observable. There is a challenge in finding a theoretical
prediction for $\left\langle v_{z}\right\rangle _{w}$ since the time states of
the quantum clocks are unknown. However, when imposing $\left(  \alpha
=0,\beta\neq0\right)  $ or $\left(  \alpha\neq0,\beta=0\right)  $, the weak
velocity becomes independent from the time states, and it follows the speed of
light with $\left\langle v_{z}\right\rangle _{w}=c$ and $\left\langle
v_{z}\right\rangle _{w}=-c$ respectively. The weak velocity of the speed of
light can also\ be found by knowing the interference between the time states,
which boils down to knowing $\tau$\ appearing\ in (\ref{v_tau1}). Following
the basic description of the one-way speed of light suggests a possible break
in the isotropy of space in the sense that the speed of light is different for
different spatial directions. The proposed setup provides a different
interpretation, where the different speeds are derived from the given pre- and
post- selected states of the quantum system. In particular, We have related
the weak velocity with the $\epsilon$ parameter in the theory of the one-way
speed of light, which allows us to re-interpret the well-known idea in which
$\epsilon$ is merely a convention assumed by an observer by its own freedom of
choice of $\epsilon$, with a physical meaning where $\epsilon$ is determined
by the weak value of the velocities governed by the pre- and post- selected
states of the system. And so the freedom to choose $\epsilon$ is converted to
the freedom of choosing the pre- and post-selected states of the system. This
connection provides a new way to link quantum mechanical behavior into
relativity theory, where instead of having a special treatment of space and
time, assuming non-isotropic space, $\epsilon$ comes from the two boundary
conditions of quantum mechanics that fully determines the velocity in between
the initial and final states of the quantum system. Thus, the proposed model
suggests that space can be, in fact, isotropic while, at the same time,
$\epsilon\neq1/2$. For future research, we propose to explore whether one can
establish a technique to synchronize the quantum clocks $A$ and $B$ by having
the detected information on the shift by the test particle. When we do not
know such synchronization, $\tau$ is, in general, unknown, and so in actual
experiments, the shift of $\Phi$ will be determined by random outcomes of
$\left\langle v_{z}\right\rangle _{w}$. Thus, one may provide a way to
statistically estimate $\tau$. We propose to explore it in future research.
For future research, we also propose to extend the model to explore the case
of measuring the speed\ of\ quantum particles that travel near strong
gravitational fields under some curved spacetime. This may provide a new way
to gain knowledge of the link between quantum mechanics and general relativity.

\bigskip

\section*{References}
[1] Reichenbach, H. (2012). The philosophy of space and time. Courier Corporation.
\newline
[2] Anderson, R., Vetharaniam, I., \& Stedman, G. E. (1998).
Conventionality of synchronisation, gauge dependence and test theories of
relativity. \textit{Physics reports,} \textbf{295}, 93-180.
\newline
[3] Blaney, T. G., Bradley, C. C., Edwards, G. J., Jolliffe, B. W., Knight, D. J. E., Rowley, W. R. C., ... \& Woods, P. T. (1974). Measurement of
the speed of light. \textit{Nature,} \textbf{251}, 46-46.
\newline
[4] Cao, S., Biesiada, M., Jackson, J., Zheng, X., Zhao, Y., \& Zhu, Z. H. (2017). Measuring the speed of light with ultra-compact radio quasars.
\textit{Journal of Cosmology and Astroparticle Physics,} \textbf{2017}, 012.
\newline
[5] Salzano, V., Dabrowski, M. P., \& Lazkoz, R. (2015).
Measuring the speed of light with Baryon Acoustic Oscillations.
\textit{Physical Review Letters,} \textbf{114}, 101304.
\newline
[6] Zhang, Y. Z. (1997). Special relativity and its experimental
foundation (Vol. 4). World Scientific.
\newline
[7] Greaves, E. D., Rodr\'{\i}guez, A. M., \& Ruiz-Camacho, J.
(2009). A one-way speed of light experiment. \textit{American Journal of
Physics,} \textbf{77}, 894-896.
\newline
[8] Finkelstein, J. (2010). Comment on \textquotedblleft A one-way
speed of light experiment\textquotedblright\ by ED Greaves, An Michel
Rodr\'{\i}guez, and J. Ruiz-Camacho [Am. J. Phys. 77 (10), 894--896 (2009)].
\textit{American Journal of Physics,} \textbf{78}, 877-877.
\newline
[9] Einstein, A. (1905). Zur elektrodynamik bewegter k\"{o}rper.
\textit{Annalen der physik,} \textbf{17}, 891-921.
\newline
[10] Page, D. N., \& Wootters, W. K. (1983). Evolution without
evolution: Dynamics described by stationary observables. \textit{Physical
Review D,} \textbf{27}, 2885.
\newline
[11] Angelo, R. M., Brunner, N., Popescu, S., Short, A. J., \&
Skrzypczyk, P. (2011). Physics within a quantum reference frame.
\textit{Journal of Physics A: Mathematical and Theoretical,} \textbf{44}, 145304.
\newline
[12] Paiva, I. L., Te'eni, A., Peled, B. Y., Cohen, E., \& Aharonov,
Y. (2022). Non-inertial quantum clock frames lead to non-Hermitian dynamics.
\textit{Communications Physics,} \textbf{5}, 298.
\newline
[13] Cohen, E. (2023). Quantum clock frames: Uncertainty relations,
non-Hermitian dynamics and nonlocality in time. \textit{Journal of Physics:
Conference Series,} \textbf{2533}. IOP Publishing.
\newline
[14] Bu\v{z}ek, V., Derka, R., \& Massar, S. (1999). Optimal quantum
clocks. \textit{Physical Review Letters,} \textbf{82}, 2207.
\newline
[15] Komar, P., Kessler, E. M., Bishof, M., Jiang, L., S$\emptyset
$rensen, A. S., Ye, J., \& Lukin, M. D. (2014). A quantum network of clocks.
\textit{Nature Physics,} \textbf{10}, 582-587.
\newline
[16] Castro-Ruiz, E., Giacomini, F., Belenchia, A., \& Brukner,
\v{C}. (2020). Quantum clocks and the temporal localisability of events in the
presence of gravitating quantum systems. \textit{Nature Communications,
}\textbf{11}, 2672.
\newline
[17] Smith, A. R., \& Ahmadi, M. (2020). Quantum clocks observe
classical and quantum time dilation. \textit{Nature communications,}
\textbf{11}, 5360.
\newline
[18] Castro-Ruiz, E., Giacomini, F., Belenchia, A., \& Brukner,
\v{C}. (2020). Quantum clocks and the temporal localisability of events in the
presence of gravitating quantum systems. \textit{Nature Communications,
}\textbf{11}, 2672.
\newline
[19] Rohrlich, D., \& Aharonov, Y. (2002). Cherenkov radiation of
superluminal particles. \textit{Physical Review A, }\textbf{66}, 042102.
\newline
[20]  Aharonov, Y., Albert, D. Z., \& Vaidman, L. (1988). How the
result of a measurement of a component of the spin of a spin-1/2 particle can
turn out to be 100. \textit{Physical Review Letters,} \textbf{60}, 1351.
\newline
[21] Aharonov, Y., \& Vaidman, L. (2008). The two-state vector
formalism: an updated review. Time in quantum mechanics, 399-447.
\newline
[22] Aharonov, Y., \& Vaidman, L. (1991). Complete description of a
quantum system at a given time. \textit{Journal of Physics A: Mathematical and
General,} \textbf{24}, 2315.
\newline
[23] Aharonov, Y., \& Shushi, T. (2022). Complex-valued classical
behavior from the correspondence limit of quantum mechanics with two boundary
conditions. \textit{Foundations of Physics,} \textbf{52}, 56.
\newline
[24] Aharonov, Y., Popescu, S., Rohrlich, D., \& Skrzypczyk, P.
(2013). Quantum cheshire cats. \textit{New Journal of Physics}, \textbf{15}, 113015.
\newline
[25] Aharonov, Y., Colombo, F., Popescu, S., Sabadini, I., Struppa,
D. C., \& Tollaksen, J. (2016). Quantum violation of the pigeonhole principle
and the nature of quantum correlations. \textit{Proceedings of the National
Academy of Sciences,} \textbf{113}, 532-535.
\newline
[26] Bliokh, K. Y., Bekshaev, A. Y., Kofman, A. G., \& Nori, F.
(2013). Photon trajectories, anomalous velocities and weak measurements: a
classical interpretation. \textit{New Journal of Physics,} \textbf{15}, 073022.
\newline
[27] Einstein, A. (1911). \"{U}ber den Einflu\ss \ der Schwerkraft
auf die Ausbreitung des Lichtes. \textit{Annalen der Physik,} \textbf{35}, 898-908.
\newline
[28] Magueijo, J. (2003). New varying speed of light theories.
\textit{Reports on Progress in Physics, }\textbf{66}, 2025.
\newline
[29] Kragh, H. S. (2006). Cosmologies with varying speed of light: A
historical perspective. \textit{Studies in History and Philosophy of Science
Part B: Studies in History and Philosophy of Modern Physics,} \textbf{37}, 726-737.
\newline
[30] Unzicker, A. (2009). A look at the abandoned contributions to
cosmology of Dirac, Sciama, and Dicke. \textit{Annalen der Physik,}
\textbf{521}, 57-70.
\newline
[31] Albrecht, A., \& Magueijo, J. (1999). Time varying speed of
light as a solution to cosmological puzzles. \textit{Physical Review D,
}\textbf{59}, 043516.
\newline
[32] Quesne, C., \& Tkachuk, V. M. (2004). Deformed algebras,
position-dependent effective masses and curved spaces: an exactly solvable
Coulomb problem. \textit{Journal of Physics A: Mathematical and General,}
\textbf{37}, 4267.
\newline
[33] Shushi, T. (2024). A Non-Standard Coupling Between Quantum
Systems Originated From Their Kinetic Energy. \textit{Annalen der Physik,}
\textbf{536}, 2300363.
\end{document}